\documentclass{cpcauth}

\usepackage{amssymb}

\begin{document}

\begin{frontmatter}



\title{From Data to Probability Densities without Histograms} 


\author{Bernd A. Berg$^{\,a,b}$ and Robert C. Harris$^{\,b}$}

\address{~~\\
$^{a)}$ School of Computational Science, Florida State University, 
Tallahassee, FL 32306-4120, USA \\
$^{b)}$ Department of Physics, Florida State University, 
Tallahassee, FL 32306-4350, USA}

\date{\today} 

\begin{abstract}
When one deals with data drawn from continuous variables, a histogram 
is often inadequate to display their probability density. It deals 
inefficiently with statistical noise, and binsizes are free parameters. 
In contrast to that, the empirical cumulative distribution function 
(obtained after sorting the data) is parameter free. But it is a 
step function, so that its differentiation does not give a smooth 
probability density. Based on Fourier series expansion and Kolmogorov 
tests, we introduce a simple method, which overcomes this problem. 
Error bars on the estimated probability density are calculated using 
a jackknife method. We give several examples and provide computer 
code reproducing them. You may want to look at the corresponding 
figures~\ref{fig_GauHist} to~\ref{fig_U1J} first.
\end{abstract}

\end{frontmatter}

\section{Introduction} \label{sec_intro}

One is often confronted with displaying an empirical probability 
density (PD) $f(x)$ of a continuous variable $x$. Most commonly this 
is done using histograms. While that is entirely appropriate and parameter 
free when $x$ is confined to discrete values, in case of a continuous 
variable $x$, one is faced with choosing a binsize, or even several 
binsizes. This is a frustrated problem: On one side one would like 
the binsize infinitesimally small, so that the resolution of the curve 
becomes perfect. On the other side the binsize has to be sufficiently 
big so that statistical fluctuations do not destroy the smoothness 
of the underlying curve. Here we do not attempt to fine-tune the 
binsize parameter(s), but propose to by-pass the entire problem by 
using a method based on the cumulative distribution function (CDF)
\begin{equation} \label{CDFdef}
  F(x)\ =\ \int_{-\infty}^x dx'f(x')\ .
\end{equation}
$F(x)$ is a monotonically increasing function as $f(x)$ cannot 
be negative. In a range with $f(x)>0$, $F(x)$ is strictly monotone.

Given a time series of $n$ real numbers (data), a parameter free 
estimate of the CDF, called empirical CDF (ECDF), is well-known 
\cite{vdW69,Be04}: The step function $\overline{F}(x)$ which, 
after sorting the data, increases by $1/n$ at each data point. 
Unfortunately, that does not help directly in getting a good estimate 
of the probability density. The derivative of a step function is a sum 
of Dirac delta functions, whereas the probability density
is often known to be a smooth function, as will be assumed in the
forthcoming. So it appears that one needs some kind of interpolation
of the CDF before taking the derivative. This is no fun, as one has
to decide whether the interpolation of 2, 3, 4, or $k$ points will 
work best. In contrast, plotting a histogram of the data is simple
and robust.  However, it is tedious to guess a smooth function from 
a histogram. Typically, either the statistical errors on the bins 
are small, but the resolution of the function is bad, or the 
resolution is good, but the statistical errors are large.

Let $\overline{F}_0(x)$ be the straight line, which is zero for
$x\le x_{\pi_1}$ and one for $x\ge x_{\pi_n}$, where $x_{\pi_1}$
is the smallest and $x_{\pi_n}$ the largest data point. After 
subtracting $\overline{F}_0(x)$ from $\overline{F}(x)$, a function 
is left over, which is zero for $x\le x_{\pi_0}$ and $x\ge x_{\pi_n}$, 
and in-between well-suited for a Fourier series expansion. This leads 
to the desired smooth approximation as long as the expansion is 
sufficiently short, but will imitate every wiggle of the data, when 
carried too far. Therefore, one needs a cut-off criterion. We provide 
this by using the Kolmogorov \cite{Ko33,vdW69,Be04} test, which tells 
us whether the difference between the ECDF and an analytical 
approximation of the CDF is explained by chance. The result is a 
well-defined, smooth empirical estimate of the PD. Our approach is 
so simple and straightforward, that one can hardly imagine that it 
is original, but we have not seen it in use before. Whether there 
exists previous literature on it or not, it is certainly desirable 
to bring it to the attention of a general science community. With 
this paper we also distribute startup software, which should help 
to bridge initial barriers against applying the approach.

In the next section we review the CDF, its empirical estimate, and 
other preliminaries. Section~\ref{PD} explains our construction 
of PDs and gives numerical examples. We consider independent events 
sampled from (1)~a Gaussian distributions, (2)~a Cauchy distribution, 
and data from (3)~a Markov chain Monte Carlo simulation, which creates 
an autocorrelated time series. Summary and conclusions follow in 
section~\ref{Conclusions}. The appendix gives a listing of the main 
routine and explains Web access to Fortran~77 code, which reproduces
the examples of this paper.

\section{Cumulative Distribution Functions} \label{CDF}

Assume we generate $n$ random numbers $x_1$, $\cdots$, $x_n$ according 
to a distribution function $F(x)$. We may re-arrange the $x_i$ in 
increasing order. Denoting the smallest value by $x_{\pi_1}$, the 
next smallest by $x_{\pi_2}$, etc., we arrive at
\begin{equation} \label{order_stat}
  x_{\pi_1} \le x_{\pi_2} \le \dots \le x_{\pi_n} 
\end{equation}
where $\pi_1,\dots , \pi_n$ is a permutation of $1,\dots ,n$. As long 
as the data are not yet created, the $x_{\pi_i}$ are random variables, 
afterwards they are data. An estimator for the distribution function 
$F(x)$ is the ECDF
\begin{equation} \label{ECDFdef}
  \overline{F} (x) = {i\over n}~~~{\rm for}~~~x_{\pi_i} \le x <
  x_{\pi_{i+1}},
\end{equation}
with $i=0, 1,\dots, n-1, n$ and the definitions $x_{\pi_0}=-\infty$,
$x_{\pi_{n+1}}=+\infty$. Confidence limits can be obtained from the 
bimodal distribution: With $x_q$ defined by $F(x_q)=q$, $0\le q\le 1$,
the probability to find $k$ data points with values $x_{\pi_1}\le\dots
\le x_{\pi_k}<x_q$ and $n-k$ data points with $x_q<x_{\pi_{k+1}}\le\dots
\le x_{\pi_n}$ is given by
\begin{equation} \label{bimodal}
  b_n(k,q) = \left(\frac{n}{k}\right) q^k (1-q)^{n-k}\ .
\end{equation}

\begin{figure}[t]
 \begin{picture}(150,155)
    \put(0, 0){\includegraphics{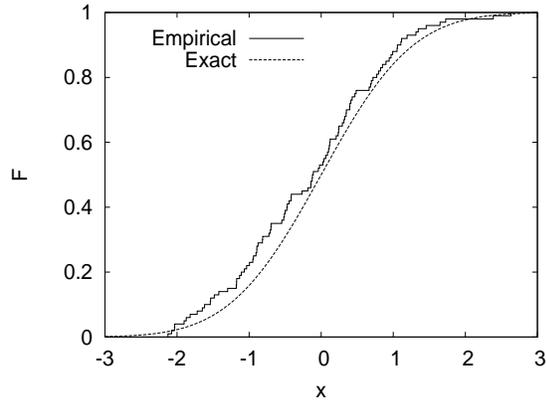}}
  \end{picture}
\caption{ECDF from 100 Gaussian random numbers versus exact 
Gaussian CDF. \label{fig_GauCDF100} }
\end{figure} 

Figure~\ref{fig_GauCDF100} shows an ECDF from 100 Gaussian distributed
random numbers together with the exact CDF. Using Marsaglia \cite{Ma90} 
random numbers, the data were created for the probability density 
\begin{equation} \label{GauPD}
  g(x)\ =\ \frac{1}{\sqrt{2\pi}}\, \exp\left(-\frac{x^2}{2}\right)
\end{equation}
and sorted with the heapsort algorithm. The CDF is in this case 
determined by the error function:
\begin{equation} \label{GauCDF}
  G(x) = \int_{-\infty}^x dx'g(x') = \frac{1}{2} + \frac{1}{2}\,
         {\rm erf}\left(\frac{x}{\sqrt{2}}\right)\ .
\end{equation}
The computer code used was that of Ref.~\cite{Be04}. 

In contrast to histograms as estimators of a PD from data, the ECDF has 
the advantage that no free parameters are involved in its definition, 
but the probability density of events is encoded in its slope. This 
makes it often impossible to read off from graphs like 
Fig.~\ref{fig_GauCDF100} high probability regions, in particular,
the point of maximum likelihood. An ECDF is well-suited for 
determining confidence intervals, however, and that can be further 
improved by switching from the CDF to the peaked CDF \cite{Be04} 
defined by
\begin{equation} \label{PCDFdef}
  F_p(x) = \cases{ F(x)\ {\rm for}\ F(x) \le {1\over 2}\,;\cr
               1 - F(x)\ {\rm for}\ F(x) > {1\over 2}\,. } 
\end{equation}
By construction the maximum of the peaked CDF $F_p(x)$ is at the median 
$x_{1\over 2}$ and $F_p(x_{1/2})=1/2$. Therefore, $F_p(x)$ has two 
advantages over the CDF $F(x)$: The median is clearly exhibited and the 
accuracy of the ordinate is doubled. It looks a bit like a PD, but it 
is in essence still the integrated PD.

\begin{figure}[t]
 \begin{picture}(150,155)
    \put(0, 0){\includegraphics{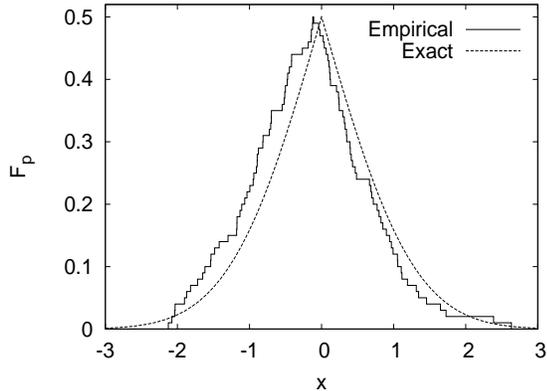}}
  \end{picture}
\caption{Peaked ECDF from 100 Gaussian random numbers versus exact 
Gaussian peaked CDF. \label{fig_GauPCDF100}  }
\end{figure} 

The peaked ECDF $\overline{F}_p(x)$ is defined in the same way, just 
replacing $F(x)$ in Eq.~(\ref{PCDFdef}) by $\overline{F}(x)$. 
Figure~\ref{fig_GauPCDF100} displays the peaked ECDF and peaked CDF 
for the same data as used in Fig.~\ref{fig_GauCDF100}. While in these
two figures deviations of the estimates from the exact function due 
to statistical deviations are clearly visible, the picture changes 
dramatically when one increases the number of data by a factor of 100. 
\begin{figure}[t]
 \begin{picture}(150,155)
    \put(0, 0){\includegraphics{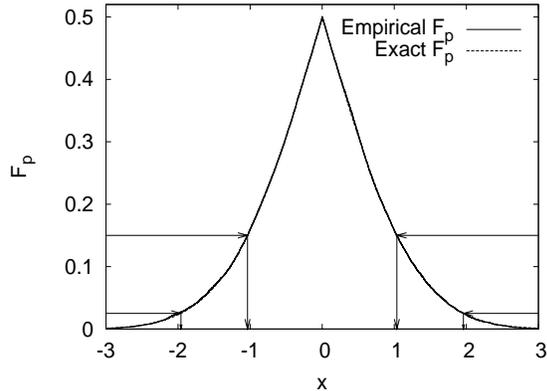}}
  \end{picture}
\caption{Peaked ECDF from 10$\,$000 Gaussian random numbers versus 
exact Gaussian peaked CDF. The arrows indicate 70\% and 95\% 
confidence intervals. \label{fig_GauPCDF10000}  }
\end{figure} 
Figure~\ref{fig_GauPCDF10000} shows the peaked ECDF from 10$\,$000 
Gaussian random numbers together with the exact peaked CDF. A difference
is no longer visible to the naked eye. Note that for large $n$ use of 
a fast sorting algorithm is mandatory. For a heapsort the CPU time 
scales with $n\,\log_2(n)$. For details see, e.g., Ref.~\cite{Be04}.

Estimation of confidence intervals is also illustrated in 
Fig.~\ref{fig_GauPCDF10000}: Just pick the desired likelihood 
on the ordinate and follow the arrows to their termination points. 
Call such a point $x_y$, then the probability for $x<x_y$ is given 
by the value on the ordinate for arrows emerging from the left, and 
the probability $x>x_y$ for arrows emerging from the right. Using one 
arrow from the left, and one arrow from the right, the probability 
content for the interval between these $x$ values is obtained by
subtracting the appropriate numbers from one (70\% for the $x$ values 
from the two inner and 95\% for the $x$ values from the two outer 
arrows of Fig.~\ref{fig_GauPCDF10000}).

Do the empirical and exact CDFs of our figures agree? The Kolmogorov
test \cite{Ko33} answers this question. It returns the probability $Q$,
that the difference between the analytical CDF and an ECDF from 
statistically independent data is due to chance. If the analytical CDF 
is exactly known and the data are indeed sampled from this distribution, 
$Q$ is a uniformly distributed random variable in the range $0<Q<1$. 
Turned around, if one is not sure about the exact CDF, or the data, 
or both, and $Q$ is small (say, $Q<10^{-6}$) one concludes that the 
difference between the proposed CDF and the data is not due to chance, 
but that one of the assumptions made is false.

Kolmogorov's ingenious test relies on no more than the maximum
difference $\triangle $ between the ECDF and the CDF. For the 
one-sided tests the relationship to $Q$ is analytically known
\cite{BiTi51,Be04} for any value of~$n$. In practice one prefers
the two-sided test for which $\triangle$ is defined by
\begin{equation} \label{KomDel}
  \triangle\ =\ \max_x\left|F(x)-\overline{F}(x)\right|\ .
\end{equation}
An exact analytical conversion into $Q$ is then not known, but an 
asymptotic expansion due to Stephens \cite{St70} gives satisfactory 
results for $n\ge 4$, which is for practically all applications 
sufficient. The test yields $Q=0.19$ for the data of 
Fig.~\ref{fig_GauCDF100} and~\ref{fig_GauPCDF100}, and $Q=0.78$ 
for the data of Fig.~\ref{fig_GauPCDF10000}. Both values are 
consistent with the assumption that the difference between the 
data and the exact CDF is due to chance.

\section{Probability Densities} \label{PD}

We would like to construct an empirical probability density (EPD) from 
an ECDF given by Eq.~(\ref{ECDFdef}). The first idea, that comes to 
mind, is to differentiate smooth interpolations of $\overline{F}(x)$.
This turns out to be tedious as long as one is unable to find a simple, 
generic rule, which determines the optimal interpolation for the 
purpose at hand, and we do not pursue this line. 

The method we propose consists of two well-defined steps.
\begin{enumerate}
\item Define as an initial approximation to $\overline{F}(x)$ a 
      differentiable, monotonically increasing function $F_0(x)$, 
      with details as specified below.
\item Fourier expand the remainder until the Kolmogorov test yields 
      $Q\ge Q_{\rm cut}=1/2$ (there may be some flexibility in 
      lowering $Q_{\rm cut}$).
\end{enumerate}
For $F_0(x)$ we require 
\begin{eqnarray} \label{F0a}
  F_0(x) &=& 0~~{\rm for}~~x\le a\,,\\ \label{F0b}
  F_0(x) &=& 1~~{\rm for}~x\ge b\,,
\end{eqnarray}
where the interval $[a,b]$ has to lie within the range of the data:
\begin{equation} \label{ab}
  x_{\pi_1}\le a < b \le x_{\pi_n}\ .
\end{equation}
For PDs with support on a compact interval, or with fast fall-off
like for a Gaussian distribution, the natural choice is $a=x_{\pi_1}$
and $b=x_{\pi_n}$. In case of slow fall-off, like for a Cauchy
distributions, or other distributions with outliers, one has to 
restrict the analysis to $[a,b]$ regions, which are sufficiently 
populated by data. This can be interpreted as considering instead 
of $f(x)$ the PD
\begin{equation} \label{fab}
  f_{ab} = \cases{ c\,f(x)~~{\rm for}~~a\le x\le b\,;\cr 
                   0~{\rm otherwise}\,.}
\end{equation}
Here the constant $c$ is defined by the normalization $\int dx\,
f_{ab}(x)=1$ and empirically obtained by the left-out probability
content of $\overline{F}(x)$, i.e., $c=n/n_{ab}$, where $n_{ab}$
is the number of data in $[a,b]$ and $n$ the total number of data.
We denote the CDF of $f_{ab}(x)$ by $F_{ab}(x)$ and its ECDF by
$\overline{F}_{ab}(x)$. After calculating from $\overline{F}_{ab}(x)$
an estimate $\overline{f}_{ab}(x)$ of $f_{ab}(x)$, the estimate of
$f(x)$ is for $x\in [a,b]$ given by $\overline{f}(x)=c^{-1}
\overline{f}_{ab}(x)$.

For $x\in [a,b]$ we restrict our choice of $F_0(x)$ in this paper to 
the straight line,
\begin{equation} \label{F0linear}
  F_0(x) = \frac{x-a}{b-a}~~{\rm for}~~a\le x\le b\,.
\end{equation}
More elaborate definitions will likely give improvements in a number
of situations (we comment on that in the conclusions), but would at
the present state just discourage applications. Our point here is
to show that good results are already obtained with the 
definition~(\ref{F0linear}).

Once $F_0(x)$ is defined, the remainder of the ECDF is given by
\begin{equation} \label{Rx}
  R(x) = \overline{F}_{ab}(x)-F_0(x)\,,
\end{equation}
which we expand into the Fourier series
\begin{equation} \label{Fourier}
  R(x)=\sum_{i=1}^m d(i)\,\sin\left(\frac{i\,\pi\,(x-a)}{b-a}\right)\ .
\end{equation}
The cosine terms are not present due to the boundary conditions
$R(a)=R(b)=0$. The Fourier coefficients follow from
\begin{equation} \label{FCof}
  d(i) = \sqrt{\frac{2}{b-a}} \int_a^b dx\,R(x)\,
         \sin\left(\frac{i\,\pi\,(x-a)}{b-a}\right)
\end{equation}
Because in our case $R(x)$ is the difference of a step function and 
a linear function, the integrals over the flat regions of the step
function are easily calculated, and the integrals (\ref{FCof}) are 
solved by adding them up.

The Fourier expansion (\ref{Rx}) is useless for a too large value
of $m$, because it will then reproduce all statistical fluctuations
of the data. To get around this problem, we perform the two-sided
Kolmogorov test first between $\overline{F}_{ab}(x)$ and $F_0(x)$ 
($m=0$), and then each time $m$ is incremented by $m\to m+1$. Once 
$Q\ge Q_{\rm cut} = 1/2$ is reached for the Kolmogorov $Q$, we know 
that the difference between the data and our analytical approximation 
is explained by statistical fluctuations. The other way round, the 
only information left in the data is statistical noise. Therefore, 
the expansion is terminated at that point. The thus obtained smooth
estimate of the CDF, 
\begin{equation} \label{Fest}
  F_{\rm estimate}(x) = F_0(x)+R(x)\,, 
\end{equation}
yields $\overline{f}_{ab}(x)$ by differentiation, and our final 
estimate of the desired PD is: 
$\overline{f}(x) = n_{ab}\overline{f}_{ab}(x)/n$.

One likes to attach error bars to the estimate of the PD. We
do this by dividing the (unsorted) original data into jackknife
\cite{Qu56,Tu58} bins and repeat the analysis for each bin. Comparing 
the function values thus obtained at selected points, error bars 
follow in the usual jackknife way (see \cite{Be04} for technical 
details).

\subsection{Gaussian distribution}

\begin{figure}[t]
 \begin{picture}(150,155)
    \put(0, 0){\includegraphics{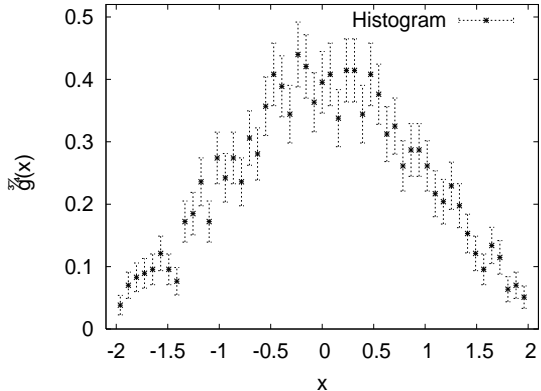}}
  \end{picture}
\caption{Histogram from 2$\,$000 Gaussian random numbers.
\label{fig_GauHist}  }
\end{figure} 

\begin{figure}[t]
 \begin{picture}(150,155)
    \put(0, 0){\includegraphics{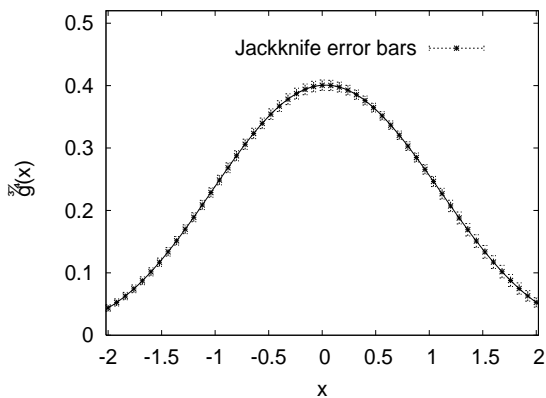}}
  \end{picture}
\caption{Estimate $\overline{g}(x)$ of the Gaussian PD from the
same data as used in Fig.~\ref{fig_GauHist}. \label{fig_GauJ}  }
\end{figure} 

Figure~\ref{fig_GauHist} shows a histogram of 51 bins for 2$\,$000 
random numbers generated according to the Gaussian distribution 
(the error bars follow follow from the variance $p\,(1-p)$ of the 
bimodal distribution~(\ref{bimodal}) with $p=h(i)/n$).

Figure~\ref{fig_GauJ} shows the estimate $\overline{g}(x)$ obtained
from the same data with the method described in this section. We used
$a=x_{\pi_1}$ and $b=x_{\pi_n}$. For the estimate from all 2$\,$000 
data, $Q=0.97$ was reached with $m=4$ ($Q=0.056$ with $m=3$). Twenty
jackknife bins were used to calculate the error bars. As all results 
of this section, the analysis is fully reproducible with the programs 
provided on the Web as described in the appendix.

\subsection{Cauchy distribution}

We consider the Cauchy distribution defined by the PD
\begin{equation} \label{CauchyPD}
  f_c(x) = {1\over \pi\,(1+x^2)} 
\end{equation}
which leads to the CDF
\begin{equation} \label{CauchyCDF}
  F_c(x) = \int_{-\infty}^x f_c(x')\,dx' = {1\over 2}+{1\over\pi}\,
  \tan^{-1} \left( x \right)\ .
\end{equation}
Due to the slow $\sim x^{-2}$ fall-off of the Cauchy PD, neither 
its mean nor its variance are defined.

\begin{figure}[t]
 \begin{picture}(150,155)
    \put(0, 0){\includegraphics{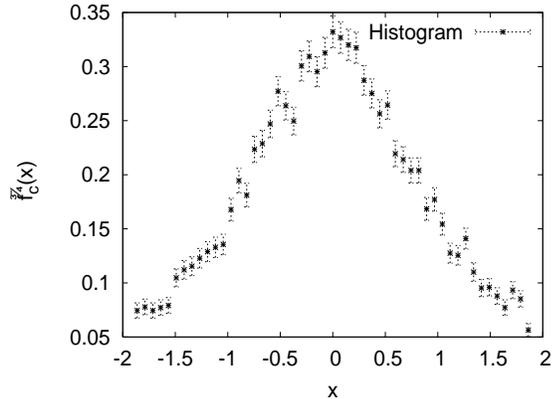}}
  \end{picture}
\caption{Histogram from 20$\,$000 Cauchy distributed random numbers.
\label{fig_CauHist}  }
\end{figure} 

\begin{figure}[t]
 \begin{picture}(150,155)
    \put(0, 0){\includegraphics{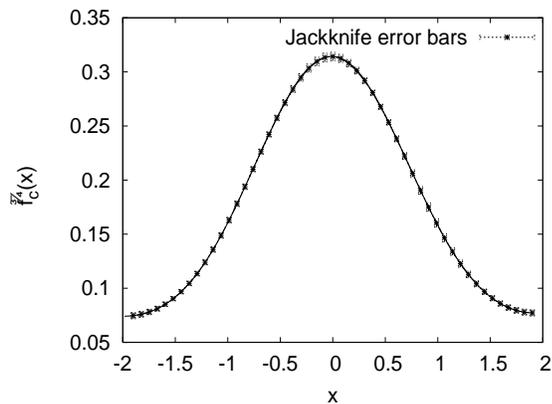}}
  \end{picture}
\caption{Estimate $\overline{f}_c(x)$ of the Cauchy PD from the
same data as used in Fig.~\ref{fig_CauHist}. \label{fig_CauJ}  }
\end{figure} 

Therefore, it comes to no surprise that we encounter considerably more
difficulties in applying our method for Cauchy data than for Gaussian 
data or for the data of our next example. To overcome instabilities, 
the analysis has to be restricted to the central region and far more 
data than before are needed for stable estimates. We ended up with 
using 20$\,$000 Cauchy distributed random number and restricting the 
analysis to
$$ a=x_{\pi_{3001}}~~{\rm and}~~b=x_{\pi_{17000}}\,,$$
which came out to be $a=-1.984$ and $b=1.916$. Figure~\ref{fig_CauHist}
depicts the 51-bins histogram of the data in this region. The estimate 
of the PD $\overline{f}_c(x)$ with our method is then shown in 
Fig.~\ref{fig_CauJ}. As before, twenty jackknife bins were used 
for the error estimates and for the estimate from all data $Q=0.77$ 
was obtained for $m=2$ ($Q<10^{-14}$ for $m=1$).

\subsection{Double peak from Markov Chain Monte Carlo}

Markov chain Monte Carlo (MCMC) simulations are widely employed 
in physics and other disciplines. Data in a MCMC time series are
autocorrelated, which makes straightforward application of the 
Kolmogorov test questionable. Here we illustrate for an example 
from lattice gauge theory (LGT) a way to deal with this problem.

In 4D U(1) LGT, a double peak in the action has been observed 
on symmetric lattices \cite{Je83}. This is characteristic for a
first order phase transition, although the situation in 4D U(1) 
LGT is somewhat questionable due to the weakness of the transition 
and other circumstances.

For this paper we have used the biased Metropolis-Heatbath algorithm
of Ref.~\cite{BaBe05} to generate U(1) data at $\beta=1.007$ on an 
$8^4$ lattice. These are parameters appropriate for producing a double 
peak. In LGT and statistical physics a standard unit for the time 
series is one ``sweep'', which corresponds for sequential updating 
(as used in our simulations) to updating each dynamical variable once. 
We have generated a time series of 2\,560$\,$000 sweeps, which takes 
about twenty hours on a 2GHz PC.

We calculated the integrated autocorrelation time $\tau_{\rm int}$ 
of our time series and found $\tau_{\rm int} \approx 1\,280$ sweeps. 
As the effective number of statistically independent data in an 
autocorrelated time series is given by \cite{Be04} $n/\tau_{\rm int}$, 
our measurement of $\tau_{\rm int}$ implies that we have generated 
approximately $2\,560\,000/1\,200 =2\,000$ independent measurements. 
In the spirit of data reduction, we then selected $2\,000$ action 
values, separated by steps of $1\,200$ sweeps, from the time series.

\begin{figure}[t]
 \begin{picture}(150,155)
    \put(0, 0){\includegraphics{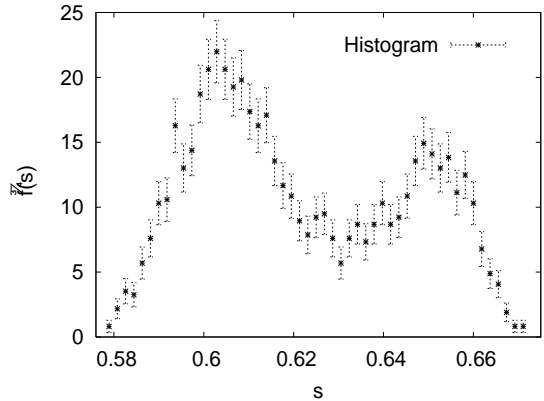}}
  \end{picture}
\caption{Histogram from 2$\,$000 effectively independent action 
measurements in U(1) lattice gauge theory. \label{fig_U1Hist}  }
\end{figure} 

\begin{figure}[t]
 \begin{picture}(150,155)
    \put(0, 0){\includegraphics{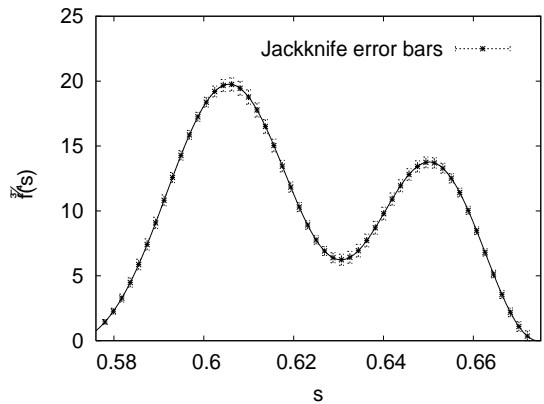}}
  \end{picture}
\caption{Estimate $\overline{f}(s)$ of the PD from the
same data as used in Fig.~\ref{fig_U1Hist}. \label{fig_U1J}  }
\end{figure} 

For these $2\,000$ measurements, the histogram with 51 entries is 
plotted in Fig.~\ref{fig_U1Hist}, where $s$ is the action density. 
Figure~\ref{fig_U1J} shows the estimate $\overline{f}(s)$ obtained 
from the same data with our
method, using $a=x_{\pi_1}$ and $b=x_{\pi_n}$. For the estimate 
from all 2$\,$000 data $Q=0.93$ was reached with $m=5$ ($Q=0.026$ 
with $m=4$), and twenty jackknife bins were used to calculate the 
error bars. The improvement from Fig.~\ref{fig_U1Hist} to
Fig.~\ref{fig_U1J} is as good as the corresponding improvement
for the Gaussian distribution.

This result sheds new light on an effect, which is well-known to workers 
in MCMC simulations, but has to our knowledge not been satisfactorily 
explained, the amazing smoothing, which one obtains when one includes 
all autocorrelated data in the histogram. In our case this is 
640$\,$000 data points, as we have taken measurements every 4 sweeps. 
The histogram error bars then fall almost on top of the jackknife error 
bars of Fig.~\ref{fig_U1J}. How can that be when there is little or no 
additional information in the autocorrelated data? Part of the 
answer appears to be that most of this information is already in 
the statistically independent data, but is usually not exploited.

\section{Summary and Conclusions} \label{Conclusions}

To someone used to plotting histogram, our method may appear difficult, 
but indeed it is not. True, first one has to sort the data to 
calculate their ECDF, then from it the empirical PD, using Fourier 
series expansion and Kolmogorov tests, and finally jackknife error 
bars need to be calculated, but all these steps are standard. Besides 
the subroutine shown in the appendix, another short subroutine, and 
the main program, we had no programming to do. All other routines were 
already available in the Web based Fortran~77 code of Ref.~\cite{Be04}. 
Once the archive described in the appendix is downloaded from the Web, 
and the program is up and running, the extra effort compared to 
plotting histograms is almost negligible. And the differences between 
Fig.~\ref{fig_GauHist} and Fig.~\ref{fig_GauJ}, Fig.~\ref{fig_CauHist} 
and Fig.~\ref{fig_CauJ}, and finally Fig.~\ref{fig_U1Hist} and 
Fig.~\ref{fig_U1J} speak for themselves.

It is astonishing that these results could be obtained with a simple
straight line (\ref{F0linear}) as initial approximation for the CDF. 
There is certainly space for improvement at the price of giving up 
the presently achieved simplicity. For our examples one and two,
the exact CDFs are known and pointed out by Eqs.~(\ref{GauCDF}) 
and (\ref{CauchyCDF}). Obviously, one does not test our method by 
constructing $F_0(x)$ from them in these examples, but they could 
be good choices, when one has reasons to expect them to be a close, 
albeit not exact, approximation of the investigated distribution. For 
instance, for the double peak of our third example, a good initial 
guess could be a sum of two Gaussians, and one could start off by 
fitting their CDF to the data. Afterwards care has to be taken that 
Eqs.~(\ref{F0a}) and~(\ref{F0b}) remain valid. That this 
can always be achieved follows from the interpretation (\ref{fab}) of 
that requirement. We abstained from investigating a double Gaussian 
$F_0(x)$, because it starts to get tedious and there is nothing really 
to improve on our result. But, one can easily imagine that there are 
more complicated situations, accompanied by limited statistics, where 
an improvement of $F_0(x)$ becomes essential. If the problem for which 
this happens is important too, it will become worthwhile to explore 
more $F_0(x)$ functions.

With our $Q_{\rm cut}=1/2$ rule, we are slightly overexpanding the 
Fourier transformation (\ref{Fourier}). In the average $Q$ should 
be 1/2, but all our final values are $\ge 1/2$. That gives some
flexibility to lower $Q_{\rm cut}$, which should be used with 
discretion in situations were the $m$ of the Fourier expansion
(\ref{Fourier}) appears to be too large. A warning right away: 
The only situation in which we did not see a rapid approach towards 
$Q> Q_{\rm cut}$ turned out to be one in which we had not noticed
that the underlying distribution was discrete, while the Kolmogorov 
test did notice that.

We did not develop a statistically rigorous approach. We address
physicists and others, who do not hesitate to use whatever works,
not those, who want to forbid numerical recipes. We rely on the 
assumption that the Fourier series (\ref{Fourier}) would be rapidly
convergent, when the ECDF in Eq.~(\ref{Rx}) would be replaced by the 
corresponding (unknown) exact CDF. That is in spirit similar to picking 
a primer in Bayesian statistics, when a rigorous one is not known, a
procedure, which can modify the probability content of confidence 
intervals.

\bigskip
\noindent {\bf Acknowledgments:}
BB likes to thank Alexei Bazavov for useful discussions. This work 
was in part supported by the U.S. Department of Energy under contract 
DE-FG02-97ER41022.

\appendix
\section{Computer Code} \label{Code}

An archive with Fortran~77 example runs can presently be downloaded 
from the website of BB. Google {\tt Bernd Berg}, or go directly to
\smallskip

\centerline{\tt http://people/scs.fsu.edu/\~\,berg/~.}
\smallskip

\noindent Take from there the research link and download the gzipped 
archive
$$ {\tt STMC\_CDFtoPD.tgz}\ . $$
Unfold the archive. (If instructions for that are needed, they can be 
found on the website of Ref.~\cite{Be04}, which is also linked on the 
main website of BB.) Go then to the folder 
$$ {\tt STMC\_CDFtoPD/Work/CDFtoPDexamples}\ . $$
All three examples of this paper can be run as special cases of the 
program {\tt cdf\_to\_pd.f} in this folder, and more instructions are 
given in its {\tt readme.txt} file. To show that at its heart our 
method is indeed quite simple, we list in the following our main 
subroutine. 
\medskip

\begin{tiny} \begin{verbatim}
      SUBROUTINE CDF_PD(IUO,NDAT,SDAT,Fxct)
C Bernd Berg, Robert Harris Dec 16 2007.
C Transforms an empirical cumulative distribution function (CDF) into 
C a corresponding probability density using and initial function plus 
C Fourier series expansion for the CDF.
C On INPUT:
C IUO     Write unit, unchanged on exit.
C NDAT    Number of input data, unchanged on exit.
C SDAT    Sorted input data, unchanged on exit.
C INTERNAL:
C Qcut    Cut-off. Fourier expansion is terminated for Q>Qcut.
C         cumulative distribution function.  
C NMAX    Maximum number of terms in the Fourier series.
C On OUTPUT:
C Fxct    Exact CDF (means here analytical approximation of the CDF).
      include '../../Libs/Fortran/implicit.sta'
      include '../../Libs/Fortran/constants.par'
      PARAMETER(Qcut=HALF,NMAX=100) ! Change also in functions.
      DIMENSION SDAT(NDAT),Fxct(NDAT)
      COMMON /CDFProb/ XMIN,XRANGE,DN(NMAX),M ! Expansion parameters.
C
      XMIN=SDAT(1) ! Initializations. 
      XRANGE=SDAT(NDAT)-SDAT(1)
      DO J=1,NDAT
        Fxct(J)=(SDAT(J)-SDAT(1))/XRANGE
      END DO
      DO K=1,NMAX
        DN(K)=ZERO
      ENDDO
      DO M=1,NMAX ! Integration for the Fourier series coefficients:
        DO I=1,NDAT-1
           X1=(SDAT(I)-SDAT(1))/XRANGE
           X2=(SDAT(I+1)-SDAT(1))/XRANGE
           DN(M)=DN(M)+(ONE*I/NDAT-X1)*COS(M*PI*X1)/(M*PI)
           DN(M)=DN(M)-(ONE*I/NDAT-X2)*COS(M*PI*X2)/(M*PI)
           DN(M)=DN(M)+SIN(M*PI*X1)/(M*M*PI*PI)
           DN(M)=DN(M)-SIN(M*PI*X2)/(M*M*PI*PI)
        ENDDO
        DN(M)=DN(M)*TWO
        DO K=1,NDAT ! CDF in Fourier series approximations:
           XRANGE1=SDAT(K)-SDAT(1)
           Fxct(K)=Fxct(K)+DN(M)*SIN(M*PI/XRANGE*XRANGE1)
        ENDDO
        CALL KOLM2_AS(NDAT,Fxct,DEL,Q) ! Kolmogorov test.
        IF(Q.GT.Qcut) GOTO 1
      ENDDO
      WRITE(IUO,'(/," CDF_PD failed M,Q =",I6,G12.3)') M,Q
      STOP "CDF_PD: Expansion failed."
1     WRITE(IUO,'(/," CDF_PD: Final M,Q =",I6,G12.3,/)') M,Q
C     
      RETURN
      END
\end{verbatim} \end{tiny} \smallskip

With exception of the jackknife routine 
$${\tt dat\_ to\_ datj.f}\,,$$
to be found in {\tt Libs/Fortran} of the package, all other 
subroutines used are from the code of Ref.~\cite{Be04} and for 
convenience included here.  For the Kolmogorov test in the form 
of Stephens the routine {\tt kolm2\_as.f} of \cite{Be04} has been 
modified to abort in case of no convergence. (Note also that the 
related routine {\tt kolm2\_as2.f} of \cite{Be04}, which is not used 
in this paper, does not work for large values of the input arguments 
$N_1$, $N_2$ due to bad coding of a multiplication of $N_1$ and 
$N_2$, which can be easily corrected.) All routines are copyrighted 
by their authors, who are listed in one of the first lines of each 
routine. Limited permission of their use is given under the conditions 
stated on the Fortran download page of Ref.~\cite{Be04}.

\end{document}